\documentclass[aip,amsmath,amssymb,reprint]{revtex4-1}

\usepackage{graphicx}
\usepackage{dcolumn}
\usepackage{bm}

\usepackage{verbatim} 
\usepackage[utf8]{inputenc}
\usepackage[T1]{fontenc}
\usepackage{mathptmx}

\usepackage{xcolor}

\usepackage{natbib}
\usepackage[colorlinks, citecolor=red]{hyperref}

\preprint{APS/123-QED}
\begin{document}

\title{Highly efficient phase-tunable photonic thermal diode} 

\author{G. Marchegiani}%
 \email{giampiero.marchegiani@nano.cnr.it}
\affiliation{
NEST Istituto Nanoscienze-CNR and Scuola Normale Superiore, I-56127 Pisa, Italy}%
\author{A. Braggio}%
\affiliation{
NEST Istituto Nanoscienze-CNR and Scuola Normale Superiore, I-56127 Pisa, Italy}%
\author{F. Giazotto}%
\affiliation{
NEST Istituto Nanoscienze-CNR and Scuola Normale Superiore, I-56127 Pisa, Italy}

\date{\today}

\begin{abstract}
We investigate the photon-mediated thermal transport between a superconducting electrode and a normal metal. When the quasiparticle contribution can be neglected, the photon-mediated channel becomes an efficient heat transport relaxation process for the superconductor at low temperatures, being larger than the intrinsic contribution due to the electron-phonon interaction. Furthermore, the superconductor-normal metal system acts as a nearly-perfect thermal diode, with a rectification factor up to $10^8$ for a realistic aluminum superconducting island. The rectification factor can be also tuned in a phase-controlled fashion through a non-galvanic coupling, realized by changing the magnetic flux piercing a superconducting quantum interference device (SQUID), which modifies the coupling impedance between the superconductor and the normal metal. The scheme can be exploited for passive cooling in superconducting quantum circuits by transferring heat toward normal metallic pads where it dissipates more efficiently or for more general thermal management purposes.
\end{abstract}

\maketitle 
The field of quantum technologies~\cite{MacFarlane2003,Riedel_2017,Acn2018} is rapidly developing thanks to the advances in nanofabrication techniques. One of the main goals is the development of quantum processors, to outperform classical computers (quantum supremacy~\cite{preskill2012,Harrow2017}). Superconducting elements constitute one of the most promising platforms for qubits implementation~\cite{DevoretMartinis2004,Wendin2017,Krantz2019}, and recently the quantum advantage has been experimentally demonstrated~\cite{Arute2019}. The requirement of a high degree of entanglement over suitable times imposes very low operating temperatures to reduce the thermal dephasing. In  state-of-the-art dilution refrigerators, the base temperature for the substrate phonons $T_{\rm ph}$ is of the order of a few mK, but the electronic temperature in the superconductors ($T_S$), which determines the environmental noise in the system, is typically somewhat higher~\cite{GiazottoRMP2006,Muhonen2012}. At the same time, the electron-phonon heat current is exponentially suppressed~\cite{TimofeevPRL102,Maisi_e-ph} $\dot Q_{\rm e-ph}\propto T_S^5\exp[-1.764 T_c/ T_S]$ when $T_S$ is much smaller than the superconducting critical temperature $T_c$. In these conditions,  electronic cooling of superconductors is an extremely complicated task since the cooling power strongly reduces for $T_S\ll T_c$. Different strategies have been proposed and exploited in thermal devices based on superconducting junctions, such as the tunnel coupling with a normal metal, which thermalizes better with the phononic environment ~\cite{GiazottoNature,FornieriReview,TimossiRouter,NguyenPRApplied2014}. This solution is not optimal for quantum computing purposes, where the galvanic coupling could interfere with the operation of the multiqubit system, introducing unwanted additional noise components. Photon-mediated thermal transport would potentially produce a smaller disturbance, since wireless coupling gives the possibility to locate the region where the power is evacuated physically far away from the superconducting device. This mechanism plays a relevant role at sufficiently low temperatures, as demonstrated both theoretically~\cite{SchmidtPRLPhotonic,PascalPRBPhotonic} and experimentally~\cite{MeschkePhotonic,TimofeevPRLPhotonic,MottonenPhotonic,PRApplied8_GrapheneJJ}, and a few prototypes of photonics-based quantum heat valves between two normal metal resistors have been recently realized, too ~\cite{RonzaniPhotonic,MailletPhotonic,SeniorPhotonic}. Heat management in low-temperature superconductors is a fundamental resource towards different applications~\cite{MarchegianiEngine,MarchegianiCooler,MarchegianiPRL,scharf2020topological}.
\begin{figure}[tp]
	\begin{centering}
		\includegraphics[width=\columnwidth]{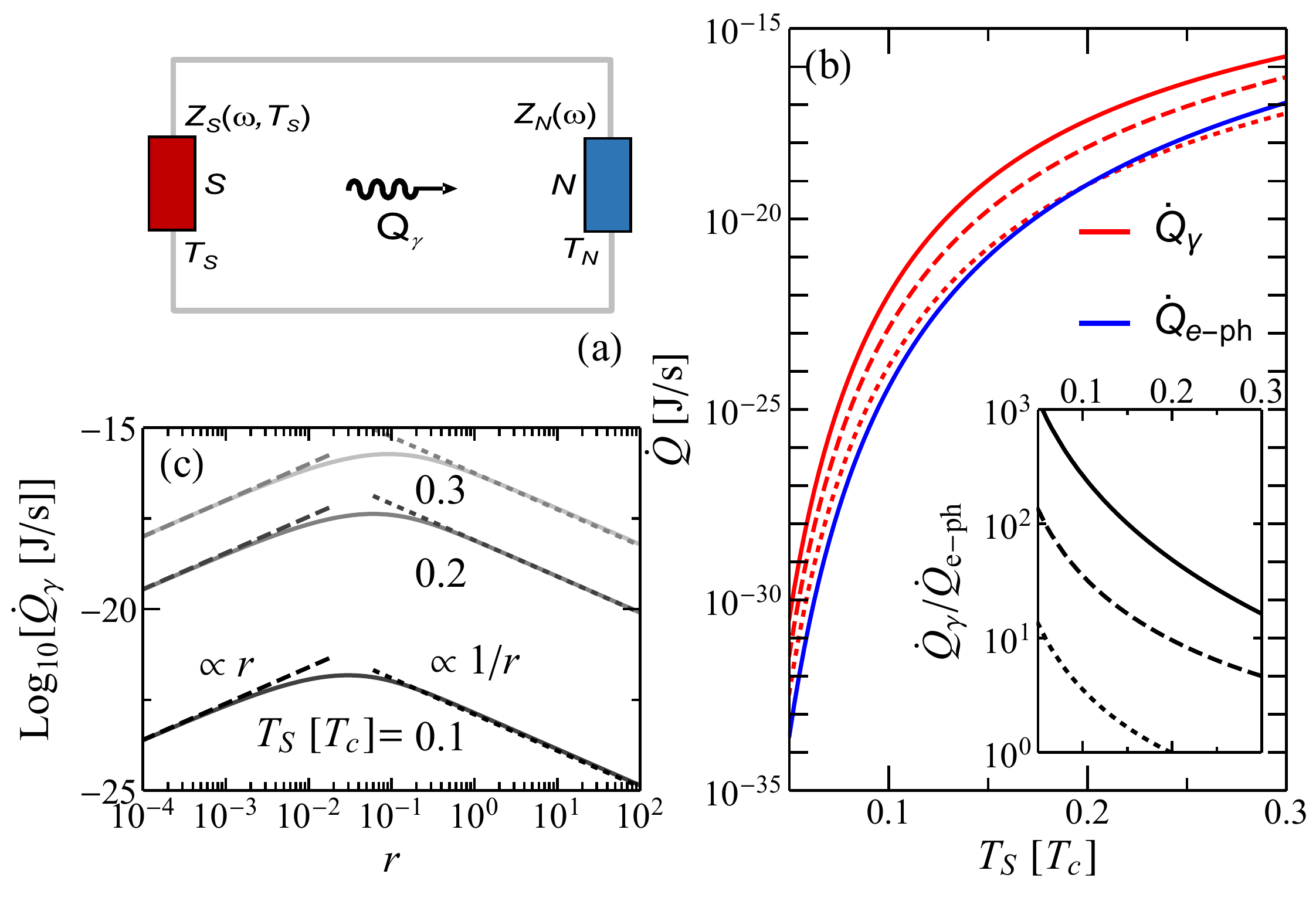}
	    \caption{\textbf{(a)} Scheme of the photonic cooler: a superconducting island ($S$) is coupled with a normal metal electrode ($N$) through a high-gap superconducting line, and the heat current is predominantly exchanged through a photon-mediated mechanism, at low temperatures.
		\textbf{(b)} Temperature evolution of the heat current removed from $S$ due to the electron-phonon exchange $\dot Q_{\rm e-ph}$ (blue) and through photon-mediated coupling $\dot Q_\gamma$ (red) with $N$ as in panel (a), for $r=R_N/R_S=0.1$ (solid), 1 (dashed), 10 (dotted). Inset: $\dot Q_\gamma/\dot Q_{\rm e-ph}$ vs $T_S$ for the same values of $r$. \textbf{(c)} $\dot Q_\gamma$ vs $r$ for different values of $T_S$. Parameters are $\mathcal V= 4\times 10^{-21} {\rm m}^3$, $\Sigma=2\times 10^8 {\rm W} {\rm m}^{-3}{\rm K}^{-5}$~\cite{GiazottoRMP2006}, $T_{c}=1.19$K, $\tau_{\rm rel}=\hbar/(20\Delta_0)$.
		}
		\label{Fig1}
	\end{centering}
\end{figure}

Here we investigate the photon-mediated thermal transport between a superconductor and a normal metal, showing that it can be more efficient than the intrinsic electron-phonon mechanism at very low temperatures. Furthermore, we demonstrate that the system acts as a very efficient thermal diode, where the heat current evacuated from the superconducting island is much larger than the opposite process. This 
envisions a passive temperature stabilization of the superconducting element.

To discuss the photon-mediated mechanism in a realistic case, we first consider the scheme in Fig.~\ref{Fig1}a. The superconducting electrode ($S$) consists of an aluminum strip of typical length $1\mu$m and cross-section $20\times200$ nm$^2$, with critical temperature $T_c^{\rm Al}=1.19$ K, and normal state dc-conductivity $\sigma_{\rm dc}^{\rm Al}=2.8\times10^7$ S/m. The superconductor is galvanically connected to a normal metal film ($N$)~\footnote{Typical materials are copper (Cu) and aluminum-manganese (AlMg)} through two wires made of a superconductor with a larger energy gap, such as niobium (Nb) or niobium-nitrate (NbN). In this setting, the superconducting wires operate as Andreev mirrors for quasiparticle thermal transport between $S$ and $N$. The photon-mediated heat transport between $S$ and $N$ reads~\cite{SchmidtPRLPhotonic,PascalPRBPhotonic}:
\begin{equation}
\dot Q_\gamma(T_S,T_N)=\int_0^{+\infty} d\omega \frac{\hbar \omega}{2\pi} \tau(\omega,T_S)[n(\omega,T_S)-n(\omega,T_N)]
\label{eq:Qgamma}
\end{equation}
where $\hbar$ is the reduced Planck constant, $\omega$ is the frequency, $T_N$ is the electronic temperature of $N$, and $n(\omega,T_i)=\{\exp[\hbar\omega/(k_B T_i)]-1\}^{-1}$ (here, $i=\{S,N\}$ and $k_B$ is the Boltzmann constant) is the Bose distribution. The photonic transfer function follows from the impedance matching between the two terminals, namely:
\begin{equation}
\tau(\omega,T_S)=4\frac{{\rm Re}[Z_S(\omega,T_S)]{\rm Re}[Z_N(\omega)]}{|Z_{\rm tot}(\omega,T_S)|^2},
\label{eq:tauphotonic}
\end{equation} where $Z_S(\omega,T_S)$ and $Z_N(\omega)$ are the impedances of the superconductor and the normal metal, respectively, and $Z_{\rm tot}(\omega,T_S)=Z_S(\omega,T_S)+Z_N(\omega)$ is the total series impedance of the circuit. In the calculations, we realistically include a finite relaxation time $\tau_{\rm rel}$ for the two electrodes. In particular, from the Drude formula, $Z_N(\omega)=R_N(1-i\omega\tau_{\rm rel})$ ($R_N$ is the dc resistance of the normal metal), and $Z_S(\omega,T_S)$ for the superconductor is obtained by extending the Mattis-Bardeen expression~\cite{MattisBardeen} of the complex conductivity to the dirty regime, as reported in Refs.~\onlinecite{ZimmermannPhysC,PrachtIEEE}. In this work, unless differently specified, we consider a dirty system with relaxation time $\tau_{\rm rel}=\hbar/(20\Delta_0)$, where $\Delta_0=1.764 k_bT_c$ is the superconducting gap, since the dirty regime represents the typical experimental situation. To consider different experimental settings, we introduce also a parameter $r=R_N/R_S$, where $R_S=Z_S(\omega=0,T_S>T_c)$ is the dc resistance of the superconductor in the normal state. When the superconductor is in the normal state, the value $r=1$ represents the impedance matching condition, with unitary transmission $\tau(\omega,T_S)=1$~\cite{Nyquist1928}. 
\begin{figure}[tp]
	\begin{centering}
		\includegraphics[width=\columnwidth]{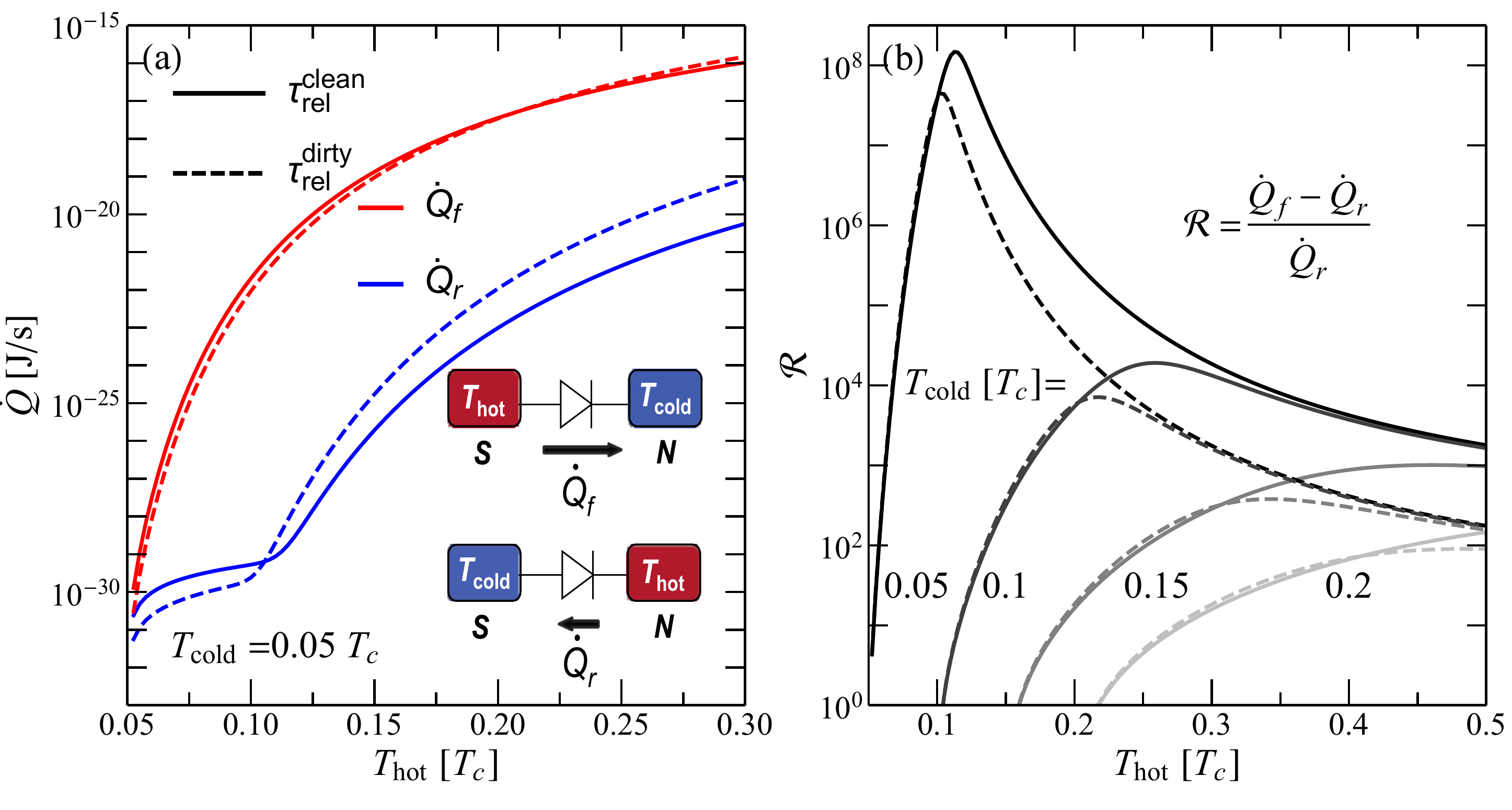}
	    \caption{\textbf{(a)} Photonic heat current vs temperature of the hot electrode $T_{\rm hot}$ for $T_{\rm cold}\to 0$. In the forward configuration $T_S=T_{\rm hot}>T_{\rm cold}=T_N$, the heat current $\dot Q_f$ is extracted from the superconductor, while in the reverse configuration $T_N=T_{\rm hot}>T_{\rm cold}=T_S$, the heat current $\dot Q_r$ is injected in the superconductor. The system acts as a thermal diode, being $\dot Q_f\gg \dot Q_r$.
		\textbf{(b)} Rectification factor $\mathcal R=(\dot Q_f-\dot Q_r)/\dot Q_r$ vs $T_{\rm hot}$ for different values of $T_{\rm cold}$. Parameters are $r= 0.05$, $\tau_{\rm rel}^{\rm clean}=5\hbar/\Delta_0$, $\tau_{\rm rel}^{\rm dirty}=\hbar/(200\Delta_0)$.
		}
		\label{Fig2}
	\end{centering}
\end{figure}

We wish now to compare the strength of the different mechanisms which lead to thermal relaxation of the superconducting island, in a situation where the temperature of the superconductor $T_S$ is larger than the environment temperature, i.e., the phononic temperature $T_{\rm ph}$ and the electronic temperature of the normal island $T_N$. Figure~\ref{Fig1}b displays the photon-mediated current $\dot Q_{\gamma}$ and the phononic heat current~\cite{TimofeevPRL102,Maisi_e-ph} $\dot Q_{\rm e-ph}\sim 0.98\Sigma \mathcal V T_S^5\exp[-1.764 T_c/ T_S]$~\footnote{In the calculation we use the integral formula of Refs.~\onlinecite{TimofeevPRL102,Maisi_e-ph}, but in the text we report the approximate expression, which substantially gives the same results in the low-temperature limit considered here.} (here $\Sigma$ is a material-dependent constant~\cite{GiazottoRMP2006} and $\mathcal V$ is the volume of $S$) flowing out of the superconductor as a function of $T_S$ for $T_N=T_{\rm ph}\to 0$ and different values of $r$. Both $\dot Q_{\rm e-ph}$ and $\dot Q_{\gamma}$ decrease very rapidly by lowering the temperature, due to the gapped excitation spectrum of the superconductor. Notably, below a certain threshold temperature, depending on the value of $r$, $\dot Q_{\gamma}$ is much larger than $\dot Q_{\rm e-ph}$. This characteristic is better visualized in the inset, where the ratio $\dot Q_{\gamma}/\dot Q_{\rm e-ph}$ is displayed in the same temperature range. In particular, the ratio increases monotonically by lowering $T_S$, and can reach a value as high as $10^3$ for $T_S=0.05 T_c$ and $r=0.1$. We can conclude that the photon-mediated channel is extremely more efficient than the intrinsic phononic channel in the heat evacuation process at sufficiently low temperatures~\cite{BosisioPhotonic}, thus the cooling of the superconducting element through the photonic coupling may appear appealing at very low temperatures. 

Note that the thermal currents are extremely low for $T_S<0.1 T_c$, being in the range $10^{-33}-10^{-23}$ J/s. However, we stress that the direct exchange of heat through the superconducting wire is several orders of magnitude smaller, even neglecting the Andreev reflection~\cite{Pannetier2000}, which further suppresses the heat flow, at the contacts between the two electrodes and the superconducting wire. In particular, the diffusive heat current can be estimated by integrating the thermal conductivity of the superconducting wire $\kappa(T)=6\Delta_w^2 \mathcal L_0\exp[-\Delta_w/(k_BT)]/(\pi^2 k_B^2 T)$, where $\Delta_w$ is the superconducting gap and $\mathcal L_0\sim2.44 \times 10^{-8}{\rm W}\cdot\Omega\cdot{\rm K^{-2}}$ is the Lorenz number. For a NbN wire ($\Delta_w\sim 2.3$meV~\cite{PrachtIEEE,DresselElectrodynamics}) of typical thickness $t_w=20$nm, length $l_w=50\mu$m, width $w_w=5\mu$m and normal state conductivity $\sigma_w=5\times 10^5$S/m~\cite{DresselElectrodynamics}, the quasiparticle diffusive heat current for $k_B T_S \ll \Delta_w$ reads~\cite{ViisanenPRBCopper,BardeenRickayzenThermalConductance} $\dot Q_{\rm wire}=\sigma_w w_w t_w/l_w\int_{T_N}^{T_S} \kappa(T) dT$. In the limit $T_N\to 0$, $\dot Q_{\rm wire}\sim 10^{-44}$J/s at $T_S=0.3 T_{c,Al}$, and exponentially suppressed at lower temperatures, hence it can be completely neglected.

Notably, the photonic heat current for $r=0.1$ is larger than for $r=1$ ($R_N=R_S$), which represents the optimal matching condition when $S$ is in the normal state. Indeed, the temperature dependence of $Z_S$~\cite{ZimmermannPhysC,DresselElectrodynamics}, yields a temperature-dependent transmission coefficient, and thereby
a non-trivial frequency matching condition. This behaviour is investigated in Fig.~\ref{Fig1}c, where $\dot Q_{\gamma}$ is plotted against $r$ for different values of $T_S$. In particular, the heat current shows a non-monotonic behavior with $r$, with a maximum for $r$ in the range 0.01-0.1, depending on $T_S$. This result can be intuitively understood as follows. At low temperatures, i.e., $T_{N},T_{S}\ll T_c$, the Bose difference in the kernel of the heat current of Eq.~\eqref{eq:Qgamma} selects the frequency with a cut-off $\hbar\omega_{cf}\sim k_B T_S\ll\Delta_0$. In this frequency range, both the real and the imaginary part of the superconductor impedance are strongly suppressed with respect to the normal state resistance~\cite{DresselElectrodynamics} due to Andreev processes $|Z_S(\omega,T_S)|\ll R_S$. As a consequence, for $r\gtrsim 1$, $Z_{\rm tot}(\omega,T_S)\sim Z_{\rm N}(\omega)\sim R_N = r R_S$, yielding a transfer function $\tau\sim {\rm Re}[Z_S]R_N/R_N^2\propto 1/r$, and hence in this case a heat current inversely proportional to $r$ (see dotted lines in Fig.~\ref{Fig1}c). However, in the limit $r\to 0$, $Z_{\rm tot}(\omega,T_S)\sim Z_{S}(\omega,T_S)$, and thus $\dot Q_\gamma,\tau\sim {\rm Re}[Z_S]R_N/|Z_S|^2\propto r$ (see dashed lines in Fig.~\ref{Fig1}c), and so the heat current admits a optimal value $r_0$ which is slightly dependent on $T_S$. 

In summary, the relevance of the photon contribution at low temperatures is generally valid for different values of $r$, and no precise tuning of the parameters is required. In the rest of this work, we set $r=0.05$: this parameter is close to the optimal condition for the low values of $T_S$ we are mainly interested in. This condition can be experimentally realized, for instance, by increasing the lateral dimensions of the normal metal electrode. This means that the volume of $N$ increases, making it even easier to keep $T_N$ at the phonons bath temperature $T_{\rm ph}$. This is beneficial for cooling purposes, since the electron-phonon interaction in the superconductor remains small, being connected to the superconductor volume, whereas the normal metal electrode is strongly coupled to the phononic bath, and consistent with the approximation $T_N=T_{\rm ph}$. 

As discussed above, the photon-mediated thermal transport represents an important contribution to the thermal relaxation of superconductors at low temperatures. Due to the asymmetric scheme, where the superconductor is coupled to a normal metal, it is relevant to inquire if there is any rectification in thermal transport. It has been reported that a junction between a normal metal and a superconductor behaves as a thermal diode~\cite{FornieriNanotechDiode}, both in the tunneling limit~\cite{FornieriAPLDiode,martinez2013efficient} and for a clean metallic contact~\cite{GiazBergNSdiode}, where the heat is transferred through quasiparticle exchange. More precisely, a thermal diode~\cite{CasatiPRL93,SegalPRL102,ROBERTS2011648,APRThermalDiodes} is a system where the current in the forward direction (in our convention, from $S$ to $N$, for $T_S=T_{\rm hot}>T_{\rm cold}=T_N$) $\dot Q_{f}=\dot Q_{\gamma}(T_S=T_{\rm hot},T_N=T_{\rm cold})$ is larger than the current in the reverse direction (from $N$ to $S$, for $T_N=T_{\rm hot}>T_{\rm cold}=T_S$), $\dot Q_{r}=-\dot Q_{\gamma}(T_S=T_{\rm cold},T_N=T_{\rm hot})$ (see the inset of Fig.~\ref{Fig2}a). 

Figure~\ref{Fig2}a displays the forward and the backward heat currents as a function of the hot temperature $T_{\rm hot}$ keeping fixed $T_{\rm cold}= 0.05T_c$, both in the clean limit regime $\tau_{\rm rel}\gg \Delta_0/\hbar$ (solid) and in the dirty limit $\tau_{\rm rel}\ll \Delta_0/\hbar$ (dashed). Note that $\dot Q_{f}$ is only slightly affected by $\tau_{\rm rel}$, whereas $\dot Q_{r}$ is typically larger in the dirty limit due to the increased exchange at subgap frequencies. Notably, in both cases, we find that $\dot Q_{f}$ is several orders of magnitude larger than $\dot Q_r$, demonstrating the generality of the thermal diode effect in the system. The physical origin of the thermal rectification effect lies in the asymmetric dependence of the transfer function Eq.~\eqref{eq:tauphotonic} with respect to the exchange of the temperature of the two terminals. Indeed, the transfer function $\tau(\omega,T_S)$ depends on the electronic temperature of the superconductor $T_S$ through the superconductor impedance $Z_S(\omega,T_S)$ but it is independent of $T_N$. Furthermore, $\dot Q_r\ll \dot Q_f$ since $\tau(\omega, T_S)\propto {\rm Re }[Z_S(\omega,T_S)]$, and $Z_S(\omega,T_S)$ is monotonically increasing with $T_S$, but strongly suppressed for $T_S\ll T_c$ such as when $S$ is colder than $N$. 

The performance of the diode effect can be quantified with a pure number, called rectification factor, here defined as $\mathcal R=(\dot Q_{f}-\dot Q_{r})/\dot Q_{r}$. This quantity is zero in the absence of rectification, i.e. for $\dot Q_{f}=\dot Q_{r}$~\footnote{The rectification is trivially zero in the absence of a temperature gradient, i.e. $T_S=T_N$, where $\dot Q_{f}=\dot Q_{r}=0$}, and can be arbitrarily large for efficient suppression of $\dot Q_{r}$. Figure~\ref{Fig2}b displays the rectification factor as a function of $T_{\rm hot}$ for different values of $T_{\rm cold}$ in the clean and in the dirty limit. For a given value of $T_{\rm cold}$, the rectification is non-monotonic with $T_{\rm hot}$, displaying a maximum (slightly dependent on $\tau_{\rm rel}$). In agreement with the discussion of Fig.~\ref{Fig2}a, the rectification is enhanced in the clean limit, due to the largest suppression of $\dot Q_r$, but the overall evolution is substantially independent on $\tau_{\rm rel}$. The highest value of the rectification $\sim 10^8$ for our parameter choice, is obtained for the smallest value of the cold temperature $T_{\rm cold}=0.05 T_c$. More precisely, the performance of the thermal diode is reduced by increasing $T_{\rm cold}$ and the rectification is zero when the superconductor is in the normal state, i.e., for $T_{\rm cold}\geq T_c $  (not shown here). Indeed, the normal-state conductance of the superconductor reads  $Z_S(\omega,T_S>T_c)=R_S(1-\omega \tau_{\rm rel})$, thus there is no rectification for $T_S>T_c$, since the transfer function $\tau(\omega)$ between two normal resistors is temperature independent.
Notably, the performance of heat rectification is still good for moderate values of $T_{\rm hot}$, with a maximum of $100$ for $T_{\rm cold}=0.2 T_c$. In particular, the latter value is of the same order of magnitude as the large rectification recently predicted in superconducting-ferromagnetic insulator tunnel junctions (where $\mathcal R\sim 500$)~\cite{GiazBergNFISdiode}, and the maximum quasiparticle rectification obtained experimentally in normal-superconductor tunnel structures ($\mathcal R\sim 140$)~\cite{FornieriNanotechDiode}. Thermal rectification has been experimentally observed also in different physical systems, including quantum dots~\cite{Scheibner_2008}, and phononics systems~\cite{KobayashiAPL95,Tian2012}, with typical rectification of order $\mathcal R\leq 1$. Moreover, our system also displays a increased performance with respect to near-field and far-field based photonic thermal diodes~\cite{OteyPRL104,AbdallahAPL,NefzaouiAPLDiode,SongAIPAdv,MirandaPhotSuper,CuevasACS2018,AbdallahAPL,OttAPL,Fiorino2018}.

\begin{figure}[tp]
	\begin{centering}
		\includegraphics[width=\columnwidth]{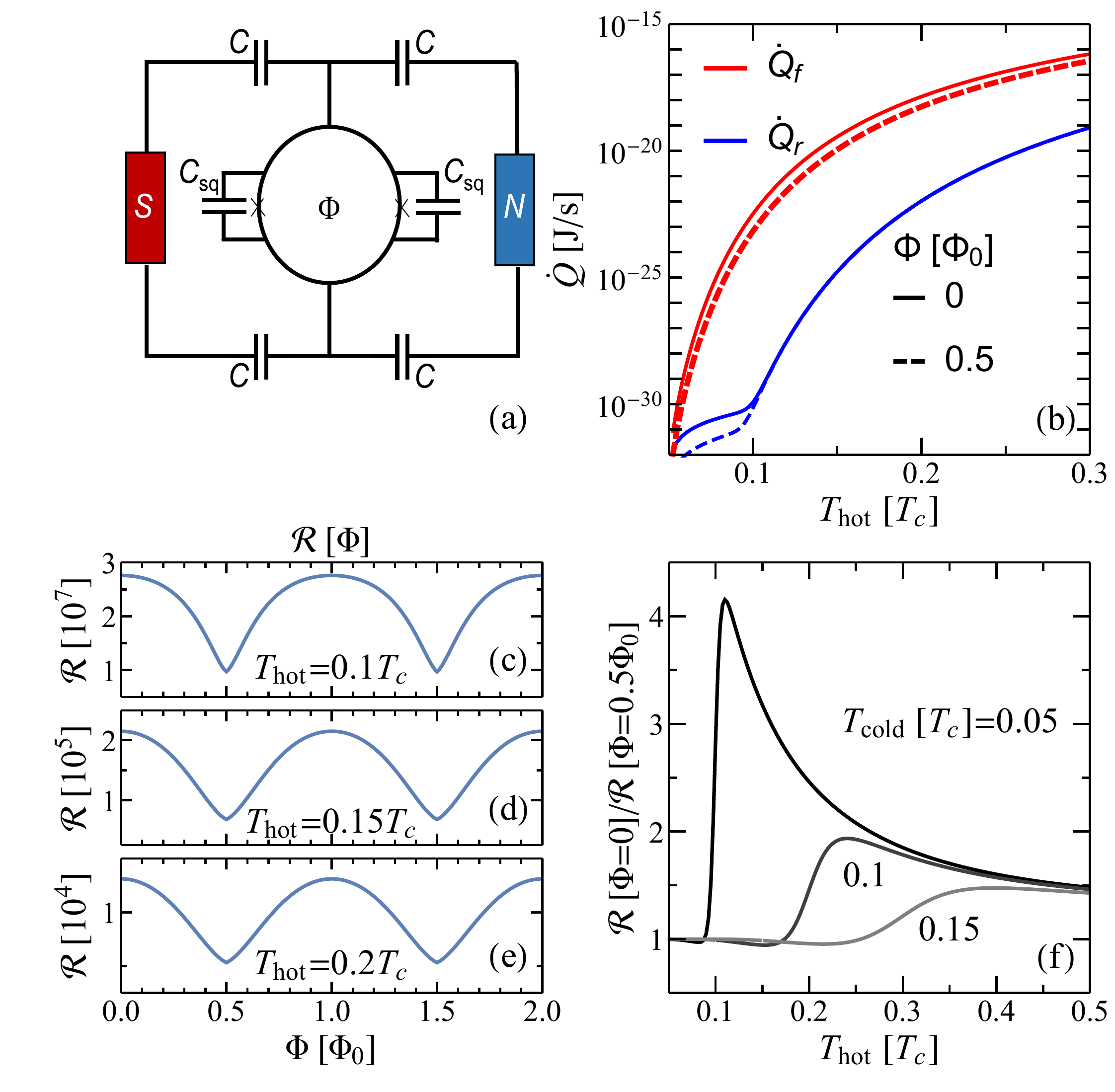}
	    \caption{\textbf{(a)} Circuit scheme of the phase-controlled photonic diode. The non-galvanic coupling is realized through series capacitors $C$. The transport is controlled thanks to a SQUID device, which behaves as an $LC$ parallel element with a phase-tunable reactive impedance $Z_{j}(\omega,\Phi)$, where $\Phi$ is the magnetic flux threading the superconducting ring. \textbf{(b)}  Photonic heat current as a function of the hot electrode $T_{\rm hot}$ for $T_{\rm cold}=0.05 T_c$ in the forward and reverse configuration. Both  $\dot Q_f$ and $\dot Q_r$ are maximum for $\Phi=0$ and slightly reduced for $\Phi=0.5\Phi_0$ 
		\textbf{(c)} Rectification factor $\mathcal R=(\dot Q_f-\dot Q_r)/\dot Q_r$ as a function of $\Phi$ for different values of $T_{\rm hot}$.	\textbf{(d)} Ratio $\mathcal R(\Phi=0)/\mathcal R(\Phi=0.5\Phi_0)$ vs $T_{\rm hot}$ for different values of $T_{\rm cold}$. Other parameters are $r= 0.05$, $I_{\rm j0}=10\mu$A, $C=50$pF, $C_{\rm sq}=25$fF, $\tau_{\rm rel}=\hbar/(20\Delta_0)$.
		}
		\label{Fig3}
	\end{centering}
\end{figure}
For practical applications, we now turn to investigate the mechanism in a non-galvanic circuit which guarantees an external tunability. This goal can be realized in different ways and a circuital optimization goes beyond the scope of this work. As a prototype example, we consider the scheme displayed in Fig.~\ref{Fig3}a, where $S$ and $N$ are galvanically disconnected through the use of capacitors $C$. The modulation of thermal transport is possible since the coupling circuit has a variable impedance $Z_{j}(\Phi)$. We envision a setup where the phase dependence of the impedance is determined by the tunable kinetic inductance of a superconducting quantum interference device (SQUID)~\cite{barone1982physics,clarke2006squid}. More precisely, the critical current of the SQUID $I_{j}(\Phi)=I_{j,0}|\cos(\pi\Phi/\Phi_0)|$ depends on the flux of the magnetic field applied externally $\Phi$, and the impedance reads~\cite{MeschkePhotonic,BosisioPhotonic,PaolucciPhotonic} 
\begin{equation}
Z_j(\omega,\Phi)=\frac{1}{i\omega2C_{\rm SQ}}\frac{\omega^2}{\omega^2-\omega_0^2(\Phi)},
\end{equation}
where $\omega_0(\Phi)=1/\sqrt{L_{\rm SQ}(\Phi)2C_{\rm SQ}}$, $C_{\rm SQ}$ is the junction capacitance of the SQUID and $L_{\rm SQ}(\Phi)=\Phi_0/[2\pi I_{j}(\Phi)]$ is the effective SQUID inductance.
In this configuration, the transmission function of Eq.~\eqref{eq:tauphotonic} depends also on the flux $\Phi$, i.e., $\tau(\omega,T_S,\Phi)$, and the effective total impedance can be now written as
\begin{align}
Z_{\rm tot}(\omega,T_S,\Phi)=&
Z_S(\omega,T_S)+Z_N(\omega)+4 Z_C(\omega)+\nonumber\\
&\frac{[Z_S(\omega,T_S)
+2Z_C(\omega)][Z_N(\omega)+2 Z_C(\omega)]}{Z_j(\omega,\Phi)},
\end{align}
where $Z_C=(i\omega C)^{-1}$ is the capacitor impedance and we used the fact that $Z_j(\omega,\Phi)$ is purely reactive to express the transfer function in the same form of Eq.~\eqref{eq:tauphotonic}.

Let us now discuss the impact of the flux-dependent coupling on the thermal diode operation of the system. Figure~\ref{Fig3}b displays the forward and the reverse heat currents as a function of $T_{\rm hot}$ for $T_{\rm cold}= 0.05 T_c$ both for $\Phi_0=0$ (solid), where the critical current is maximum, and for $\Phi_0=0.5$ (dashed), where it is zero, giving a SQUID impedance of $Z_j(\omega)=(i\omega 2C_{\rm SQ})^{-1}$. Note that the zero-flux value of $\dot Q_{f}$ is larger than the current for $\Phi=0.5\Phi_0$. Conversely, $\dot Q_{r}$ is independent on the magnetic flux for $T_{\rm hot}>0.1 T_c$, while at very low temperature $T_{\rm hot}<0.1 T_c$ it results $\dot Q_{r}(\Phi=0)>\dot Q_{r}(\Phi=0.5\Phi_0)$. As a consequence, the rectification factor can be tuned with the flux of the external magnetic field. This possibility is shown in Fig.~\ref{Fig3}c, where the rectification factor $\mathcal R$  is displayed as a function of $\Phi$, and different values of $T_{\rm hot}$. In particular, the rectification is an even and periodic function of $\Phi$ with period $\Phi_0$: it is maximum for $\Phi=k\Phi_0$ (with $k\in\mathbb Z$) and monotonically decreasing in the semi-period $\Phi\in[k\Phi_0,(k+1/2)\Phi_0]$. Moreover, the rectification is maximum at for $T_{\rm hot}\sim 0.1 T_c$, and decreases by several orders of magnitudes at larger temperatures, similar to the behaviour investigated in Fig.~\ref{Fig2}b for $T_{\rm cold}=0.05 T_c$. To give a more complete characterization of the flux control over the rectification factor, we consider the ratio $\mathcal R(\Phi=0)/\mathcal R(\Phi=0.5\Phi_0)$ as a function of $T_{\rm hot}$ for different values of the bath temperature $T_{\rm cold}$, as displayed in Fig.~\ref{Fig3}d. In particular, the rectification ratio evolves non-monotonically and it has a maximum for a given value of $T_{\rm hot}$, depending on $T_{\rm cold}$. For $T_{\rm cold}=0.05 T_c$, the zero-flux value is more than 4 times larger than the minimum value, and thus amenable to a clear experimental observation of the phase-dependent modulation.

In summary, we discussed the photon-mediated thermal transport between a BCS superconductor and a normal metal. We demonstrated that the photonic contribution plays an important role at very low temperatures. We discussed the operation of the system as a thermal diode, where the heat is evacuated in the hot superconductor, and it is strongly suppressed when the superconductor is cold. We analyzed two configurations: i) a \textit{galvanic} circuit where the superconductor is connected to the normal metal through a superconducting wire, but quasiparticle diffusion is suppressed, ii) a \textit{non-galvanic} circuit where the impedance matching can be controlled by applying a magnetic field to a SQUID. In both cases, the system acts as an extremely efficient thermal diode, with a maximum rectification factor of the order of $10^7-10^8$ at $T_{\rm hot}=0.1 T_c$. In the phase-tunable scheme, the rectification factor can be controlled and reduced up to a factor 4. This work gives a general discussion on the potential use of photon-mediated coupling for passive cooling of superconducting circuits at low temperatures. Our results can be generalized to different schemes and geometries. For instance, the cooling power could be conveniently increased by using multiple circuital lines which evacuate the heat in normal metal dissipators remotely located.

The authors acknowledge the EU's Horizon 2020 research and innovation programme under grant agreement No. 800923 (SUPERTED) for partial financial support. A.B. acknowledges the CNR-CONICET cooperation program "Energy conversion in quantum nanoscale hybrid devices", the SNS-WIS joint lab QUANTRA funded by the Italian Ministry of Foreign Affairs and International Cooperation, and the Royal Society through the International Exchanges between the UK and Italy (Grant No. IEC R2 192166). F.G. acknowledges the European Research Council under the EU’s Horizon 2020 Grant Agreement No. 899315-TERASEC for partial financial support.

%

\end{document}